# REALIZING BONE-MASS GENERATION THROUGH A DENSITY TYPE THEORETICAL ARCHETYPE

**Nirmalendu Hui** and **Biplab Chattopadhyay**[*]

Department of Physics, Acharya B. N. Seal College, J. N. Road, Cooch Behar 736101, W.B., INDIA

### Abstract

The dynamic process of the formation of bone-mass in case of humans is studied to gather precise understanding about the same with the motivation of applying these concepts for healing of bone-fracture and non-unions in non-invasive manner. Three cellular ingredients, osteoblasts, osteoclasts and osteocytes are predominant players in generating new bone-mass in which a host of hormones, proteins and minerals have potential supportive role. Considering population density of these three biological cells osteoblasts, osteoclasts and osteocytes as variables, we formulate a theoretical model, in the form of a set of time differential equations in order to emulate the dynamic process of bone-mass creation. High relative abundance of osteocytes at asymptotic scale together with moderate level values of osteoblasts and osteoclasts cell populations signifies formation of bone-matter in our theoretical archetype. The archetype has been studied through both analytical and numerical channels, significant results are emphasized and relevant conclusions are drawn. Certain predictive statements are also made which could be put to future in-vivo clinical test or in-vitro physiological experimentation as well.

**Key words:** Bone-mass; Osteoblasts osteoclasts & osteocytes; Population density model; Controlled bone formation; Clinical trial & in-vitro test.

## I. INTRODUCTION

The term Electromagnetic field (EMF) [1] represents a characteristic physical entity having its origin to the area of electricity-magnetism within the subject matter of physics. The fact that such a physical entity could be linked and be used within the domain of biology has been perceived by mankind, probably, centuries ago. The very perception has been made reality with the advent of transduction of certain types of energy in case of biological entities and the eventual functional response [2-4] of these biological entities to such transduction. As a result a new inter-phasing area of research evolved which is termed as bio-electromagnetics [5-9] in which attention is put to study the interaction between electromagnetic fields and biological systems or entities.

Bone is a biological entity observed in the animal world and the same, in aggregate, constitute the basic skeletal structure of quite a many species of animals predominantly pertaining to the mammalian group. In a word, bone is an integral and essential part of mammalian species including that of homo-sapiens. The happening of bone fracture in human physique is very common and it generates considerable predicaments, more so, when the bone fractures are of grave magnitude. Repairing or healing of such bone fractures and other related disorders such as non-unions have been done, from time immemorial, by various classical methods which had always been painful and very often invasive to a large extent. However, achievement in the direction of producing low frequency EMFs opened the pathway of applying such fields for the repairing or healing of bone fractures, non-unions and allied ailments [5-9]. The most remarkable aspect of applying EMF in case of fracture healing has been that the whole procedure is completely non-invasive, painless and mostly free from any other adverse post-healing effects.

Up till now it is emphasized that EMF in various forms could be used for fracture repair in bone and for healing other allied problems. However our understanding about the exact mechanism, through which such repair or healing takes place, is truly marginal. This, probably, is the reason why the non-invasive procedure of EMF-fracture healing is not so popular even at this age of ours. Thus, a precise understanding of the mentioned mechanism of fracture healing of bone by EMF is genuinely a necessity. It is to be noted here that to gather thorough understanding about the process of fracture healing by EMF would logically lead us to know the constitution of bone and the process through which bone-mass is being generated. Thus, with the motivation to generate command on application of EMF for fracture healing, we, as a maiden step endeavour to realize the dynamical features involved in bone-mass growth and hence this communication.

Bones in human body that make up to the skeletal structure are neither completely solid nor some unchanging entities as they are commonly perceived. In the contrary, bones do undergo the process of constant reshaping through a complex process of formation and remodelling [10, 11]. The significant cellular constituents involved in the whole process of formation and remodelling of bones are found to be osteoblast, osteoclast and osteocyte [12, 13]. Some hormones, which act as directors of protypical cellular functions, as well as some essential minerals such as calcium, phosphorous do also play significant role in the whole process of bone formation and remodelling. Out of these cellular constituents, osteoblasts are considered to be builders which also produce collagen and hydroxyapatite [4, 14]. Whereas osteoclasts, which are longer in size than osteoblasts, are observed to dissolve bone by acting on its mineral matrix part [15, 16]. These clast cells give rise to collagenese that breaks down collagen and also secretes various acids that dissolve the hydroxyapatite structure. Some of the blast cells get confined within the bone matrix and such cells are referred as osteocytes [17, 18]. Thus, osteoblasts are involved in the formation of bone and osteoclasts inflict bone resorption [19]. Normally the state of mature bone is close to an equilibrium situation where the processes of formation and resorption are in force in intertwined manner and are of comparable magnitudes keeping the required balance of bone-mass. In other words, the process of remodelling of bone does never get stopped or it is a dynamical process.

As it turns out, the whole process of bone formation is a dynamic affair where predominantly the cellular constituents osteoblast, osteoclast and osteocyte play their respective roles. It may not be out of place to mention here that the inherent dynamicity embedded within the process of bone formation, which is also termed as remodelling process, is mostly due to two reasons. Firstly, the dynamicity or remodelling in bones is necessary for adaptation to the changes inflicted under various types of loads. And secondly, continuous remodelling (dynamicity) is necessary for the repair of recurrent damages caused at the micro level of bones.

It is to be noted here that literature allied to bone-growth contains earlier theoretical work on remodelling of bone [20] emphasizing the critical role played by osteocytes. In our work significance of three different biological cells, namely osteoblasts, osteoclasts and osteocytes are elaborated and the role of their mutual interactions, on the process of bone-growth in general, has been explored. Consequently, in this paper, we are inclined to exploit the inherent dynamicity of the process of bone formation and transcribe the whole process of bone-mass generation in terms of a dynamical model involving the three significant cellular constituent osteoblasts, osteoclasts and osteocytes to be the model variables. Our motto is to gather precise understanding of the whole dynamical process of bone-growth through both the formation and resorption routes with an eye to predict and evolve potential procedures to heal or repair any externally caused damages to the bone. Such damages could be of various types such as fractures of bone, non-unions and other allied disorders.

Having narrated the motivation for taking up studies on the dynamics of bone-mass generation in the introductory section, we are to bring up the specifics of organization of the rest of this paper. In section II, we discuss about the theoretical archetype representing the dynamical process of bone-mass generation and shape the set of time-differential equations simile of the process. In section III, we make a critical analysis of the archetype in the theoretical avenue emphasizing stability of model solutions and consequently the sustainability of the archetype. Detailed numerical analysis of the archetype is done in section IV and the outcomes are elaborated with relevant planar viewgraphs. General results from our calculations are discussed in section V and logical conclusions are drawn along with some predictions.

## II. THEORETICAL ARCHETYPE OF BONE FORMATION DYNAMICS

Accumulating data from clinical research as well as experimentation involving various mammalian species, it is observed that the bone formation dynamics predominantly revolves around the interplay between three different types of cells namely osteoblasts, osteoclasts and osteocytes. It has also been emphasized in the literature that the relative population densities of the cells signify their effectiveness in the bone formation or bone degradation processes. Out of three predominant cells osteoblasts population is regarded to be responsible for formation, deposition [4] and mineralisation [19] of bone tissue. The osteoclasts population do the job of resoption at the bone [21, 22] edges to prepare the bone arena for further growth at which actually the blast cells start acting for bone formation through deposition as well as mineralisation. Towards completion of the formation process, blast cells get settled within the mineralized scaffold and are thence termed as osteocytes [18, 23]. In other words, when osteoblasts, through definitive processes, gets into the shape of osteocytes they become localized within the relevant part of the bone matter [24]. Thus abundant osteocyte population [25] could be considered as a signature of production of bone matter that eventually implies an active bone formation process to be in place at the relevant location. Further, it has been observed that during the process of bone formation the relative abundance of osteocyte population becomes more than 10 times as compared to osteoblast population [26]. It should be noted here that though osteoclastic cells act in dissolving the bone matter, their usefulness cannot be undermined. Because, while there is a fracture in bone, formation of new bone matter within the fractured gap could take place only when the fracture-edges are dissolved. In this sense osteoclast population could be regarded as precursor to the new bone formation.

Analysis of the clinical and other experimental data pertaining to the progression of bone formation, as embedded in the allied literature, thus points towards the fact that in bone mass generation three different biological cells play pivotal roles which are osteoblasts, osteoclasts and osteocytes. Taking cue from such assertions, we formulate a theoretical archetype to mimic the bone formation dynamics involving the densities of these cells' populations to be archetype variables as osteoblasts (B), osteoclasts (C) and osteocytes (S). Results from clinical and other experiments also reveal definitive processes through which blasts, clasts and cytes population densities at the relevant spatial location are being modified with progressing time starting from the initiation of the bone mass generation.

Osteoblasts population is enriched or degraded predominantly by three different processes:
(a) Blast cells which are specialised stromal cells, get differentiated from mesenchymal cells at a constant rate ($p$) depending on the activation of specific transcription factors and proteins [27].
(b) Blast cells get proliferated at a constant rate ($\alpha$) owing to their interaction with a specific set of cytokines [28].
(c) Osteoblastic cells are in constant interaction with the osteoclastic cells and thus they give vent to enhancing osteoclastic cell population by way of stimulation [15] and they also get transformed adding to the population of osteocytes. Thus, eventually blast-clast interaction inflicts a loss in blast population at a constant rate $\gamma$.

Processes regulating the osteoclastic population are as follows:
(a) Clast cells are predominantly differentiated from monocytes and macrophages [29] at a constant rate $q$.
(b) Interaction of clast population with the blast population where blast cells actually inflicts stimulation on the clast population [28-30] in which an essential enhancement of clast cells is being effected at a rate $\eta$.
(c) Clast cells undergo apoptosis [31] at a constant average rate $\mu$.

The population of osteocytes, which is considered to signify the gravity of bone formation, is being influenced by three different processes:
(a) Blast cells, being interacted by clast population, effectively get transformed into osteocytes [18, 24]. A definite fraction $\lambda$ of the interacting blast-clast populations adds to the osteocyte population.

(b) Clast cells always get them to adhere to the bone edge or bone surface and being interacted by surface sitting osteocytes, they produce resoption of the bone matter [32, 33]. This consequently means that interaction of clast cells with osteocytes gives rise to a loss of osteocytes at a constant rate $\nu$.

(c) Osteocytes are removed from the relevant spatial location by way of natural loss [34] at a constant rate $\mu'$.

Having emphasized various processes through which the balance between osteoblast, osteoclast and osteocyte populations are maintained at the relevant spatial region proximity to bone formation, we are in a position to jot-down a set of time differential equations which would be representative theoretical archetype of the bone mass generation or bone growth dynamics.

We consider three model variables such as $B, C$ and $S$ representing densities of osteoblast, osteoclast and osteocytes population (in units of $mm^{-3}$) respectively at the proximal spatial region where new bone formation is at place. To arrive at theoretical archetype corresponding to bone growth dynamics we make following assumptions:

(A1) An upstream of blast cells, near to the space of bone formation, at a constant rate $p$ is assumed. These blast cells are mostly derived from the mesenchymal cells in which transcription factors Runx and Osterix play essential role.

(A2) Osteoblast population gets proliferated owing to the action of a family of Bone Morphogenic Protiens (BMP) which are a part of Transforming Growth Factor Beta (TGF − β) super family of cytokines. We assume proliferation of blast cells by a magnitude $\alpha$ ($\in R_+$).

(A3) Blast population, owing to the interaction with clast population, makes way for enrichment of other two variables $C$ (clast cells) and $S$ (osteocyte cells) and hence suffer a loss, which is assumed to be at a constant rate $\gamma$ ($\in R_+$).

(A4) An upstream activation of clast cells is assumed to be in place at the relevant proximal space of bone formation at a constant rate $q$. Such osteoclast population is actually differentiated from haematopoietic cell lines of macrophage/monocyte linage.

(A5) Osteoblast population acting on the monocytes (precursor to clast cells) increases the proportion of osteoclast population. In our model, we improvise this to be an interactive stimulation of clast population by that of blasts and assume that such interactive activation adds to the abundance of osteoclast cells at a constant rate $\eta$ ($\in R_+$). This has also been termed in the literature as clastogenesis.

(A6) A per capita natural loss (apoptosis) of osteoclast population is assumed at a constant rate $\mu$ ($\in R_+$).

(A7) Combined actions of clast and blast cells lead to the maturation of the process of bone formation which is identified with the generation of osteocytes. These combined effects of osteoblasts and osteoclasts could be taken as an interaction between these two populations. Thus it is assumed that blast-clast interaction leads to production osteocytes at a constant rate $\lambda$ ($\in R_+$).

(A8) Osteoclast populations, acting on the bone surface, dissolve the osteocytes population embedded within the bone matrix. In biological terms, the osteoclast cells interact with the bone surface osteocytes and thus the dissolution. We assume removal of osteocytes owing to the osteoclast-osteocyte interaction at a constant rate $\nu$ ($\in R_+$).

(A9) We assume a per capita natural loss of osteocytes, which may be linked to the self renewal as well as remodelling of bone-tissues, at a constant rate $\mu'$ ($\in R_+$).

Taking into account the assumptions A1–A9, as elaborated above, involving various interactive processes among the osteoblast, osteoclast and osteocyte populations, we have the theoretical archetype elucidating the dynamical trail of bone mass generation as

$$\frac{dB}{dt} = p + \alpha B - \gamma BC$$
$$\frac{dC}{dt} = q + \eta BC - \mu C \qquad (1)$$
$$\frac{dS}{dt} = \lambda BC - \nu CS - \mu' S$$

The above theoretical model could be studied through analytical as well as computational avenues to extract different characteristic signatures of the biological process of the growth of bone and to make relevant predictions.

## III. ANALYTICAL OVERLAY AND STABILITY OF MODEL SOLUTIONS

A thorough scrutiny of archetype equations (1) reveal that all of these are smooth functions of the variables B, C and S as well as the set of parameters involved. While arriving at the theoretical archetype emulating the dynamic biological process of bone formation, we have emphasized that the three model variables and all the model parameters would remain confined on the positive domain of real line. This, in turn, assures that model solutions do exist and they hold uniqueness as well as continuity properties in the positive octant of the coordinate space.

Stability of asymptotic solutions of the model equations could be judged by analysing various equilibria of the model. Existence of well defined equilibria manifestly point towards inherent stability of the biological system or process under consideration [35]. Three different equilibria involving the asymptotic stable solutions of model variables are possible: $E_1(B_1^*, 0, 0)$, $E_2(B_2^*, C_2^*, 0)$ and $E^*(B^*, C^*, S^*)$. The equilibria could be obtained by solving the following model equations,

$$\begin{aligned} p + \alpha B^* - \gamma B^* C^* &= 0 \\ q + \eta B^* C^* - \mu C^* &= 0 \\ \lambda B^* C^* - \nu C^* S^* - \mu' S^* &= 0 \end{aligned} \quad (2)$$

In addition to these three, another trivial equilibrium $E^*(0,0,0)$ exists but this trivial equilibrium bears only marginal or no significance in terms of the biological process. This is because asymptotic stable values of all three variables representing osteoblast, osteoclast and osteocyte populations are zero here, whereas, in practical, the relative abundance of asymptotic stable populations actually determines various characteristics of the allied biological process. Solutions obtained for different nontrivial stable equilibria are as follows:

(i) With reference to $E_1(B_1^*, 0, 0)$ we obtain stable solutions for $B_1^* = \left|\frac{p}{\alpha}\right|$, where modulus is incorporated to eradicate the unphysical nature of the solution owing to a lurking negative sign.

(ii) In terms of equilibrium $E_2(B_2^*, C_2^*, 0)$ stable asymptotic variables turn out to be

$$B_2^* = \frac{(\alpha\mu - \gamma q - \eta p) \pm \{(\alpha\mu - \gamma q - \eta p)^2 + 4\alpha\eta p\mu\}^{1/2}}{2\alpha\eta} \quad \text{and}$$

$$C_2^* = \frac{q}{\mu - \eta B_2^*} \quad (3)$$

(iii) In terms of the general stable equilibrium $E^*(B^*, C^*, S^*)$ the asymptotic stable values are obtained by solving the full set of equations (2) which yields

$$S^* = \frac{\lambda B^* C^*}{\nu C^* + \mu'}, \quad (4)$$

$$C^* = \frac{q}{\mu - \eta B^*} \quad (5)$$

Whereas $B^*$ could be obtained from the quadratic equation

$$\alpha\eta B^{*2} - (\alpha\mu - \gamma q - \eta p)B^* - p\mu = 0 \quad (6)$$

Solutions of equation (6) produces

$$B^* = \frac{(\alpha\mu - \gamma q - \eta p) \pm \{(\alpha\mu - \gamma q - \eta p)^2 + 4\alpha\eta p\mu\}^{1/2}}{2\alpha\eta} \quad (7)$$

$B^*$ would assume real values confined on the positive part of real-line provided

$$(\alpha\mu - \gamma q - \eta p)^2 + 4\alpha\eta p\mu \geq 0 \quad (8)$$

The non-trivial equilibrium $E^*(B^*, C^*, S^*)$ is characteristic of the stability of asymptotic solutions of all three variables in particular and the allied biological system in general, and this characteristic general equilibrium is bound by the conditional relations involving various model parameters as depicted in equation (8). This $E^*$ is also termed as the interior equilibrium.

Stability of the general equilibrium could further be confided by satisfying Routh-Hurwitz criterion [36] for the interior equilibrium which necessitates considering the Jacobian matrix for the theoretical archetype as depicted in equations (1). For getting into Jacobian matrix, we need to linearize model equation (1) around the interior equilibrium point by incorporating a new set of variables. These new variables

signify measures of allowed deviation about the interior equilibrium point $(B^*, C^*, S^*)$. Thus we define

$$X(t) = B(t) - B^*$$
$$Y(t) = C(t) - C^* \quad (9)$$
$$Z(t) = S(t) - S^*$$

Substituting these in the model equations (1), we linearize them about the non-trivial general equilibrium point $E^*$. Following simplifications and pursuing standard established procedures, we arrive at the matrix equation for the new variables as

$$\begin{pmatrix} \dot{X} \\ \dot{Y} \\ \dot{Z} \end{pmatrix} = \begin{pmatrix} \alpha - \gamma C^* & -\gamma B^* & 0 \\ \eta C^* & \eta B^* - \mu & 0 \\ \lambda C^* & \lambda B^* - \nu S^* & -(\nu C^* + \mu') \end{pmatrix} \begin{pmatrix} X \\ Y \\ Z \end{pmatrix}$$
$$= J \begin{pmatrix} X \\ Y \\ Z \end{pmatrix} \quad (10)$$

Where J is like a transformation matrix called the Jacobian Matrix.

Our interest lies in the matter of judging the stability of general equilibrium point $E^*(B^*, C^*, S^*)$ and the same can be done by checking whether the Jacobean matrix satisfies the Routh-Hurwitz criterion which demands that

$$Tr\, J < 0 \quad and \quad det\, J > 0 \quad (11)$$

On analysing the Jacobian matrix, we obtain the trace and determinant of the same as functions of asymptotic stable values of variables and the various model parameters, which are given as follows

$$Tr\, J = (\alpha - \mu - \mu') + \eta B^* - (\gamma + \nu)C^* \quad (12)$$

$$det\, J = \alpha\mu\nu C^* - \alpha\eta\mu' B^* + \alpha\mu\mu' - \mu\mu'\gamma C^* - (\alpha\nu\eta - \mu'\gamma\mu + \mu'\eta\gamma)B^*C^* - \gamma\mu\nu C^{*2} + \gamma\eta(1-\nu)B^*C^{*2} \quad (13)$$

We have checked that the Jacobean matrix corresponding to the theoretical archetype of bone formation process satisfies Routh-Hurwitz criterion for a default set of parameters (in Table I) of the model. It could also be said that adherence to Routh-Hurwitz criterion imposes certain numerical bounds on the parameters of the model. In any case, satisfying Routh-Hurwitz criterion by the Jacobian matrix signifies that the model under considerations does have an interior equilibrium point $E^*$ and the same emphasizes that the system is locally asymptotically stable around the interior general equilibrium $E^*(B^*, C^*, S^*)$.

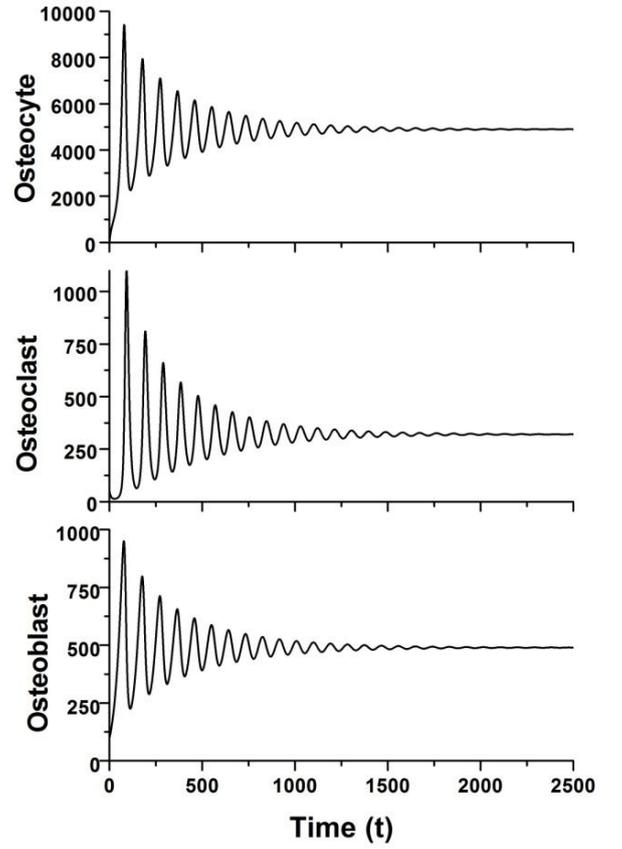

FIG. 1. Time series solutions are plotted for three model variables as labelled in the figure as functions of time in hours. Parameters are assigned their default values as in Table I.

## IV. ANALYSIS OF THE ARCHETYPE THROUGH NUMERICAL PATHWAY

Set of equations representing mathematical archetype of bone formation dynamics could be solved by standard numerical tools. These solutions provide useful information regarding characteristic features of the model as well as the patterns of bone formation dynamics. Data from numerical solutions could be analyzed graphically to understand detailed behaviour of the model and to obtain various thresholds related to any criticality existing within the model.

Archetype equations (1) are solved numerically by applying fourth order Runge-Kutta method [37]. It is to be noted that solutions of model equations require defining default or standard values of the parameters occurring in the set of equations. Default values of the parameters are preliminarily estimated by looking at signatures and strength of various interactions in the existing literatures. And these parameter values are further justified and sharpened by judging solutions of model variables such that time-variation of these solutions emulate established characteristics of any theoretical archetype corresponding to a biological system. Default values of the parameters are shown in Table I with units of each of them elaborated. Initial seed values of model variables are taken as $B(t = 0) = 100$, $C(t = 0) = 50$ and $S(t = 0) = 20$. However, it has been checked thoroughly that specifically the asymptotic solutions, or, in general the solutions do not depend on the initial values of model variables. Time variable in the archetype is taken in discrete steps of 0.001 for the purpose of numerical solution. It is imperative to mention here that, although we have defined default values of the parameters in our numerical calculations we always vary the parameters around their default values to explore any significant characteristics of the biological system under consideration. Solutions are analysed in terms of view graphs plotted in two-dimensional co-ordinate space. Two different categories of numerical data are generated. The first category being the solutions of all the model variables as functions of time and in the second category, asymptotic stable solutions are obtained as functions of various model parameters whose view-graphs could be termed as phase-curves.

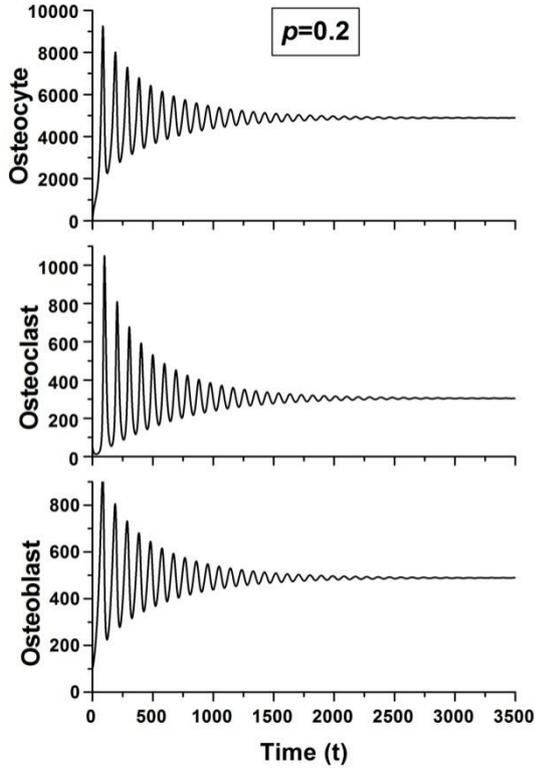

FIG. 2. Solutions for archetype variables are plotted as functions of time (in hours) with parameter $p$ set to the value as shown in the figure and all other parameters being at their default values as in Table I.

Solutions for three different model variables osteoblast, osteoclast and osteocyte populations as a function of time (taken in units of hours) is plotted in Figure 1 with the values of parameters set to their default level as given in Table I. We find that all three populations show oscillations

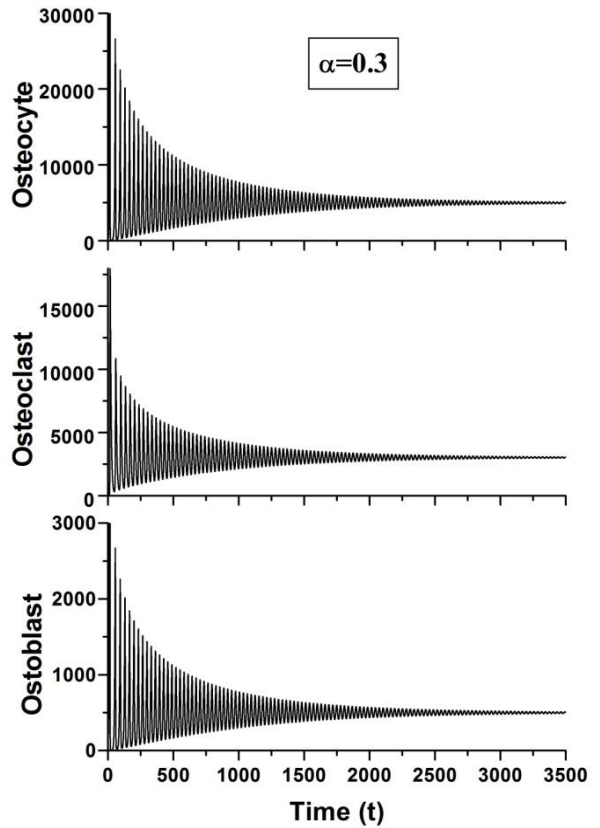

FIG. 3. Model variables Osteoblast, Osteoclast and Osteocyte populations are plotted as functions of time (in hours) for the parameter α as in figure with all other parameters being at their default values as in at initial time, amplitude of oscillations get diminished with the passage of time and all three solutions ultimately get stabilized to respective constant values at large or asymptotic time scale. Such characteristic signatures of model variable solutions are typical of any theoretical archetype of a biological system and the same emboldens the poise of modelling the bone formations dynamics. Further, the outcome in the present case, that asymptotic stable solution of osteocyte population is approximately 10 times that of osteoblast population, matches with the clinical estimation in similar situation as being observed in the literature [26]. In the process of formation of bone, osteoblast cells, when get trapped within the bone matrix through a complex process involving various interactions and stimulations among the three biological cells and many other biological ingredients, they are termed as osteocytes. In some sense, localized osteoblasts are actually osteocytes and such osteocytification [24] of osteoblasts at large time signifies maturation of the process of bone-formation. It may be noted here that the process of bone formation is carried on continuously till required amount of bone-mass is created at any specific location and such creation takes place in small amounts (or thin layers) at definite time intervals. Here, the mentioned time-interval is assimilated with that required for osteoblast, osteoclast and osteocyte populations to get stabilized and the amount of bone-mass creation in that interval is measured by the asymptotic stable numerical value of osteocyte solution. Looking at Figure-1, we can assert that in a time-interval of nearly 2000 hours (≈83 days) a moderate layer of bone-mass, comprising nearly 5000 osteocytes (in units of $mm^{-3}$) being embedded within the bone matrix, is created. Thus, in general, abundance of osteocyte cells at the asymptotic time-scale could be taken as a measure of the formation of bone-mass. Hence, larger the values of asymptotic stable solution of osteocyte population, bigger would be the magnitude of bone-mass creation and smallness of stable osteocyte value ($S^*$) would mean a situation where marginal or practically no formation of bone-mass takes place. In Figure 2, we have plotted model variables B, C and S as a functions of time with a change in the value of parameter $p = 0.2$, keeping all other parameters at their standard values as in Table-1. The parameter $p$ represents the rate of differentiation of blast cells from mesenchymal cells, and in the plot we find that the archetype is marginally sensitive to the change in parameter. Time-dynamic progression of solutions remains almost similar for the present case with $p = 0.2$ as in the standard case with $p = 1.0$ (see Figure 1). The only mismatch observed is in the extent of time-scale at which solutions become stable and single valued. In the present case ($p = 0.2$) oscillations of the solutions carried till a greater value of time ($\approx$ 2800 hours), but the asymptotic stable value of various cells (model variables) remain nearly at the same level as that for $p$ =1.0 (default value).

**Table-1:** Set of Parameters used with their default values

| Parameter | Definition | Default Value (UNIT) |
|---|---|---|
| p | Rate of upstream activation of osteoblasts | 1.0 $(mm^{-3} hour^{-1})$ |
| α | Rate of proliferation of osteoblasts owing to various cytokines | 0.03 $(hour^{-1})$ |
| γ | Removal rate of osteoblasts owing to their giving into other cells | 0.0001 $(mm^3 hour^{-1})$ |
| q | Rate of upstream activation of osteoclasts | 1.0 $(mm^{-3} hour^{-1})$ |
| η | Rate of production of osteoclasts due to blast-clast interaction | 0.0003 $(mm^3 hour^{-1})$ |
| μ | Rate of natural dissolution of osteoclasts | 0.15 $(hour^{-1})$ |
| λ | Rate of production of osteocytes due to blast-clast interaction | 0.02 $(mm^3 hour^{-1})$ |
| ν | Removal rate of osteocytes owing to the osteoclast-osteocyte interaction | 0.002 $(mm^3 hour^{-1})$ |
| $μ'$ | Rate of natural loss of osteocytes | 0.0002 $(hour^{-1})$ |

Time varying solutions for the model, are obtained for the parameter $\alpha = 0.3$ with all other parameters as defined in Table I and the same is represented in Figure 3. Here we observe that increase in the parameter $\alpha$, representing lysis-type proliferation [28] rate of blast cells owing to the action of cytokines, makes the solutions to oscillate with higher frequency. Asymptotic stable values of all model variables also find nearly 10-fold increment than their respective values for $\alpha = 0.03$, the estimated default value of $\alpha$.

In Figure 4 we plot time varying solutions for osteoblast, osteoclast and osteocyte populations for $q = 5.0$ where $q$ represents the rate of differentiation of clast cells from monocytes/ macrophages. Here we find that an increase in $q$ inflicts considerable lowering of oscillations of the model variable solutions with stabilization of solutions occurring at a much lower time scale ($\sim 500$ hours) than that for the default $q = 1.0$ (see Figure 1). Small decreases in asymptotic stable values of all variables are observed in the present case as compared to those with default set of parameters. We have also checked the plots of variables B, C, and S as function of time for increased value of $\gamma = 0.001$ than its default value as in Table I.

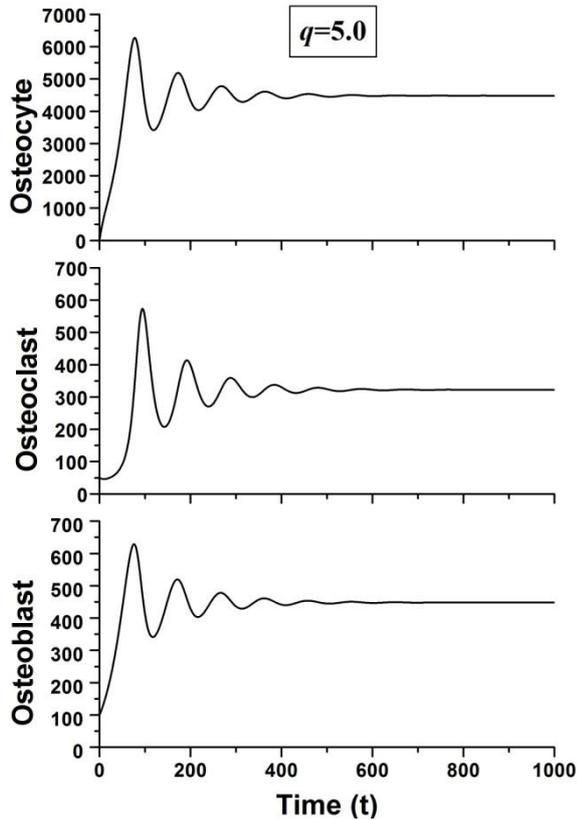

FIG. 4. Model variable solutions are plotted with increasing time (in hours) with the parameters being at their respective default values except the parameter $q$ which is assigned the value as captioned.

We observe that oscillatory characteristics of solutions die-down in a very short time and stabilization of solutions reach very fast in the time scale with asymptotic values of solutions being slightly less than the default case. Similarly we also checked plots of variables with time for a much lower value of $\eta = 0.000003$ than the estimated default value $\eta = 0.0003$. Here we find that, oscillations sustain till a large time and the asymptotic $B^*, C^*$ and $S^*$ jumps to nearly 100 times their values for default situation (as in Figure 1).

In Figure 5 we have plotted osteoblast, osteoclast and osteocyte populations as a function of time for $\mu = 1.0$, which is higher than its default value. We observe that increase in $\mu$, which gives the apoptosis rate of osteoclast cell population, makes the oscillations in the solutions to be more and time taken for such solutions to be stable single-valued get considerably increased. We also find a nearly seven fold increase in numerical value of the asymptotic stable solutions for all the variables than those for default values of all the parameters.

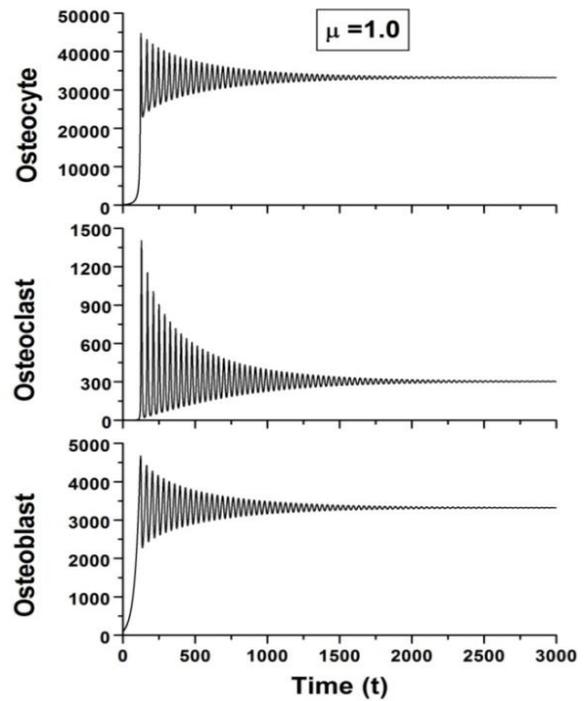

FIG. 5. Solutions for variables are plotted as functions of time (in hours) with $\mu$ as in figure and other parameters at their default values as in Table I.

Within our model, the parameter $\lambda$ represents the rate at which osteocytes are produced from osteoblast cells owing to their interaction with the osteoclast population. This process could also be termed as osteocytification of osteoblast by way of getting localized. In Figure 6 we plot model variables for $\lambda = 0.002$ which is less than its default value with all other parameters remaining at their default measure. Here we find that decrease of $\lambda$ can not inflict any significant change in the characteristic of time series solutions of archetype variables i.e., osteoblast, osteoclast and osteocyte populations.

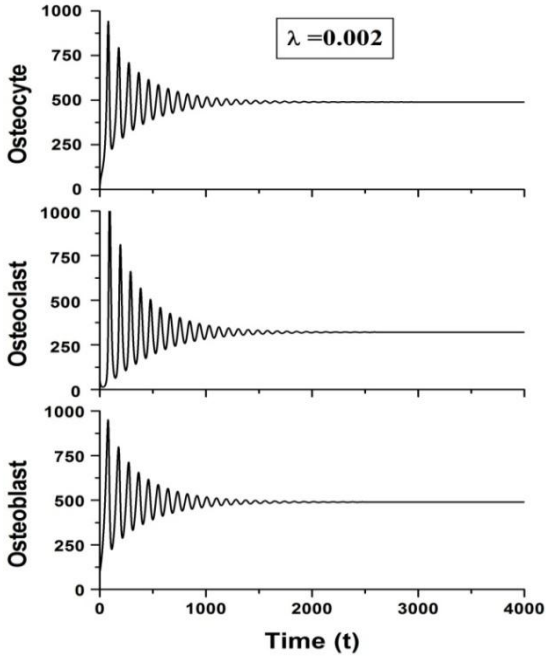

FIG. 6. Plot of archetype variables as functions of time (in hours) with λ as shown in the figure and other parameters being at their standard values as in Table I.

We have also checked that the time series solutions of archetype variables with values of $\nu$ and $\mu'$ away from their defined default values. A change in $\nu$, the rate of bone resorption by clast cells, by two orders of magnitude (setting $\nu = 0.2$) inflicts changes in the asymptotic stable values of osteocyte population. We find that although asymptotic stable values of osteoblast and osteoclast populations remain at their default levels, osteocyte population falls to two orders of magnitude. A change in $\mu'$ by nearly three orders of magnitude cannot introduce any significant change in the time series solutions of archetype variables.

Next, we move on to analyze various phase diagrams associating the archetype. As emphasized earlier, in a phase diagram, we plot asymptotic stable solutions of model variables $B^*, C^*$ and $S^*$ as functions of relevant model parameters. We have generated phase diagrams involving all the model parameters and analyzed them to bring out characteristics associate with dynamics of bone-mass generation. We also make certain predictions for future test based on the outcomes of data analysis of such phase diagrams.

Phase diagram in terms of parameter $p$ is presented in Figure 7 in vertical stack of panels. We find that at small $p$, stable asymptotic solutions for osteoblast and osteocytes exhibit quadratic increase with small amplitude oscillations around the average quadratic curve. With the magnitude of $p$ increasing, both amplitude and frequency of oscillations diminish and beyond a value of $p = 20$, curves acquire the form of average quadratic curve. At large $p$, these two populations $B^*$ and $S^*$ also monotonically saturates. Osteoclast stable solutions ($C^*$) linearly increase with the increase in $p$ and numerically it is at the similar level as $B^*$. We also notice that stable solutions of all these three populations start at non-zero finite values even with $p$ tending to zero with the proportion of $B^*$ to $S^*$ keeping at $\frac{B^*}{S^*} = \frac{1}{10}$, as emphasized in the literature [26].

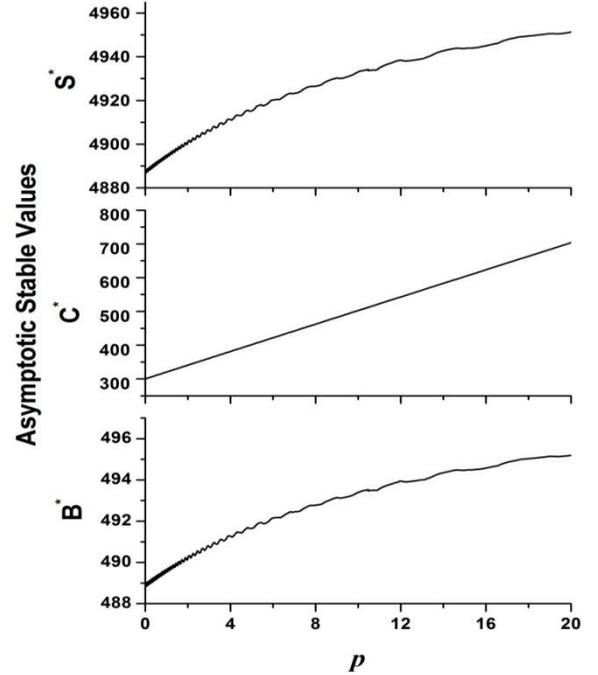

FIG. 7. Phase diagram depicting asymptotic stable values of archetype variables as functions of changing $p$ with all other parameters being kept at their default values as in Table I.

In Figure 8 we plot stable asymptotic solutions $B^*, C^*$ and $S^*$ as functions of increasing $\alpha$, the rate of proliferation of osteoblasts owing to the action of various cytokines. We find that, at around $\alpha \sim 0.5$, osteoblast and osteocyte populations both get saturated to their global stable values (for default set of parameters) keeping their $\frac{1}{10}$ proportion [26]. However osteoclast population $C^*$ remains at nearly zero for $\alpha$ tending to zero and increases linearly with the increase of $\alpha$. Relative abundance of osteoclasts, even at any moderate levels of $\alpha$ is much higher than the other two populations.

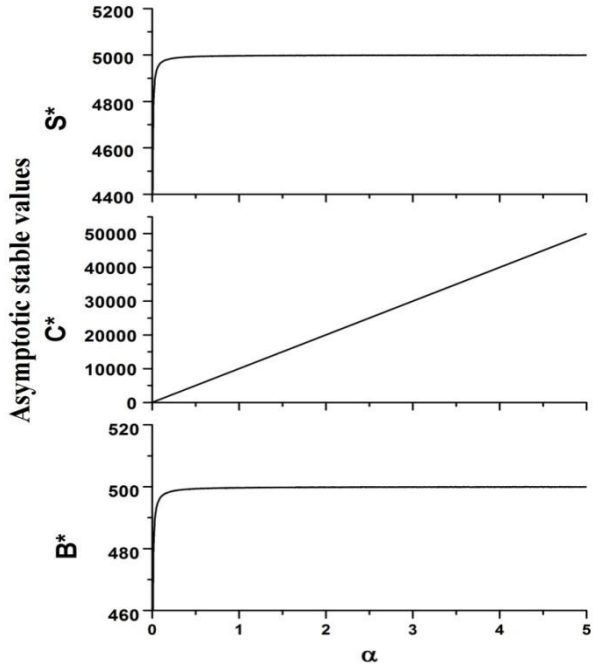

FIG. 8. Stable values of osteoblast, osteoclast and osteocyte populations are depicted in planar graph with increasing α. Other parameters are at their respective standard values as in Table I.

Asymptotic stable solutions $B^*, C^*$ and $S^*$ are plotted as functions of the parameter $q$ representing the rate of differentiation of clast cells from monocytes or macrophages, as in Figure 9. We find that increase of clast differentiation rate inflicts an oscillatory increase in the asymptotic clast population which sheds off oscillations at large $q \gtrsim 20$ but still gets enhanced steadily. However osteoblast and osteocyte asymptotic solutions ($C^*$ and $S^*$) decrease steadily and linearly with the increase of $q$ and for very large $q$ they are driven to zero.

In Figure 10 we plot $B^*, C^*$ and $S^*$ as functions of the parameter $\mu$, the apoptosis rate of osteoclast cells. Here we find that osteoblast and osteoclast solutions linearly increase, starting from zero, with the increasing magnitude of $\mu$, keeping their relative $\frac{1}{10}$ proportion intact. Also, the osteoclast asymptotic solutions turns out to be at high values for very small $\mu$, gets sharply decreased for $\mu \lesssim 0.1$ and beyond $\mu \approx 0.1$ it becomes globally steady at around $C^* \sim 300$.

Figure 11 is a plot of $B^*, C^*$ and $S^*$ versus $\lambda$, that represents generation rate of osteocytes owing the blast-clast interactions. In this case, asymptotic stable solutions $B^*$ and $C^*$ are always at their global stable values at $B^* \sim 490$ and $C^* \sim 320$, whereas, osteocyte population $S^*$ increases linearly and steadily starting at zero for $\lambda$ very small nearing to zero.

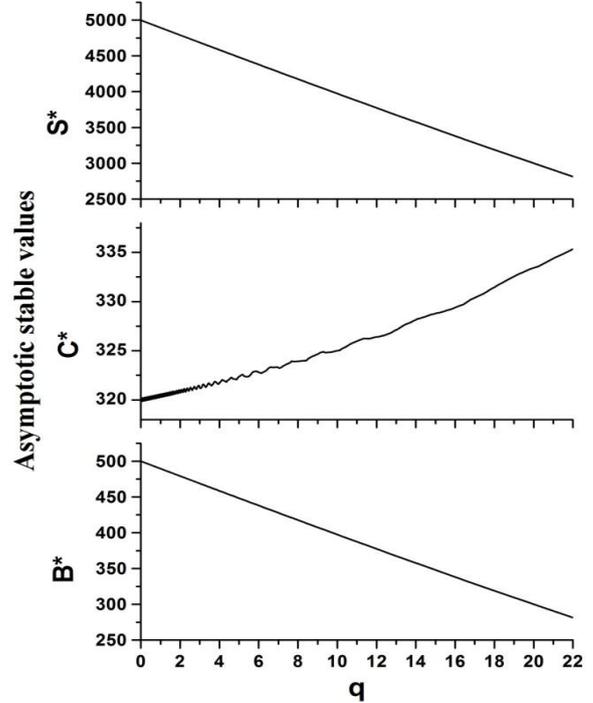

FIG. 9. Asymptotic stable archetype variables are plotted as functions of $q$ with other parameters being as in Table I.

We have also checked phase diagrams involving $\gamma$, the removal rate of blast population owing to their interaction with clast population that adds to the clast population as well as produces osteocytes. We observed that, all three parameters $B^*, C^*$ and $S^*$ are driven to zero with $\gamma$ assuming even very small value ($\sim$ 0.02). A plot of $B^*, C^*$ and $S^*$ versus $\eta$ shows that $B^*$ and $S^*$ become zero very fast ($\eta$ close to zero) and $C^*$ increases linearly at fast pace starting at zero for $\eta \sim 0$. Phase diagram with parameter $\nu$ shows that asymptotic solutions of blast and clast populations remain at their respective global stable values $B^* \sim 490$ and $C^* \sim 320$, whereas $S^*$ falls to zero even for negligibly small $\nu$. Plot of $B^*, C^*$ and $S^*$ as functions of $\mu'$ shows that $B^*$ and $C^*$ are again at their respective global stable values as mentioned earlier and $S^*$ diminishes nearly linearly starting at a value $\sim 10 B^*$ for $\mu'$ close to zero.

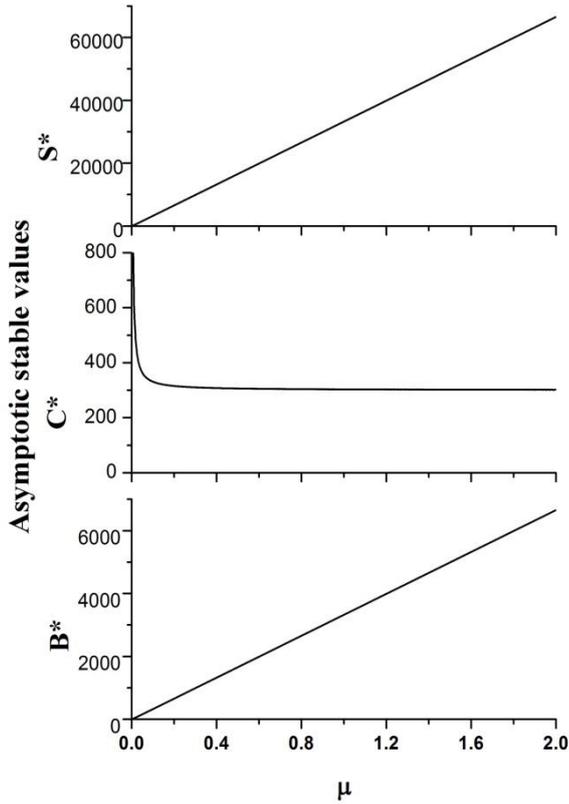

FIG. 10. Phase diagram for archetype variables where stable populations are plotted as functions of μ and with all other parameters as in Table I.

## V. DISCUSSIONS AND CONCLUSIONS

We have been elucidating the dynamical process of bone-matter formation in order to gather precise understanding about the same and with an eye to ultimately formulate non-invasive methods of bone-fracture healing by applying suitable EM fields.

Gathering various characteristic signatures of bone-formation process from the accumulated literature, we have formulated a theoretical archetype of the dynamical process of bone-formation, analyzed the same through standard analytical tools to gain confidence on the validity of the model and further did thorough numerical analysis of the archetype or model with the aid of planer view graphs. Results of our calculations are represented elaborately in the earlier section. Here we attempt to discuss the significant parts of our results, draw relevant conclusions and make certain predictions which would be put to test through clinical trials or experimentation.

Theoretical analysis of archetype reveals that there exists clearly two stable equilibria $E_2(B_2^*, C_2^*, 0)$ and $E^*(B^*, C^*, S^*)$ signifying clear sustainability of the proposed archetype of the biological process of bone-formation. The partial equilibria $E_1(B_1^*, 0, 0)$ does not lead to clean solution and may be ignored while judging the sustainability of the archetype as compared to the general equilibrium, which yields well-defined and meaningful solutions. Solutions pertaining to general equilibrium produce imposing conditions on the parameters of archetype which is obeyed thoroughly by the default set of parameters as in Table I. Further we checked that the general equilibrium fully satisfies Routh-Hurwitz criterion affirming the stability and sustainability of the archetype.

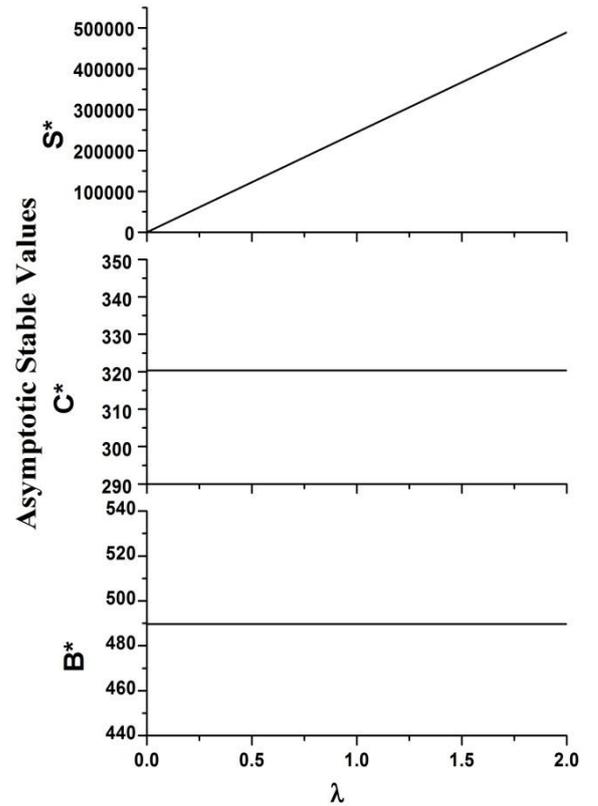

FIG. 11. Stable values of three populations (archetype variables) are depicted with the parameter l being changed starting from near zero. All other parameters are set to their default values as in Table I

Time series solutions for three different variables of the archetype, $B, C$ and $S$ representing osteoblast, osteoclast and osteocyte populations respectively, exhibit high amplitude oscillations at small times with the amplitude diminishing gradually and at asymptotic time scale all the solutions terminate to their respective stable single values. In other words, all the variables become independent of time at the asymptotic scale by virtue of their time differentials becoming zero. This is prototypical of theoretical archetype of any biological system having inherent dynamicity embedded within it. Change in the values of various parameters from their default level only inflicts changes in the magnitudes of asymptotic stable values of variables as well as shifts the cap on the time scale beyond which it is truly asymptotic. One other characteristic feature of bone-mass generation to be mentioned is that, a fixed relative balance is maintained between osteoblast and osteocyte populations ( $\frac{B^*}{S^*} = \frac{1}{10}$ ratio) for default values of model parameters as well as for the changed magnitude of parameters where ever there is signature of active bone-mass formation with abundant $S^*$.

Detailed analyses carried on various phase diagrams reveal several important conclusions regarding the favourable situations under which bone-mass formation can be boosted. The phase diagram in terms of parameter $p$ shows that an increase in the generation of osteoblasts (achieved through differentiation from mesenchymal cells) makes both $B^*$ (asymptotic stable value of osteoblast) and $S^*$ (asymptotic stable value of osteocyte) increase slowly and at large $p$ acquire their respective saturation values $B^*{\sim}500$ and $S^*{\sim}5000$. Notice that $\frac{B^*}{S^*} = \frac{1}{10}$ proportion is maintained all through. In the process of bone formation, osteoclast cells are importantly relevant, because these cells cause resorption of bone-edge and only such resorption could initiate formation of new bone-mass at larger scale. Thus, osteoclast cells are precursor to generation of new bone-mass. In other words, under any favourable situation of enhanced bone-mass generation, stable osteoclast population ($C^*$) must assume non-zero values which must be at least at the moderate level, otherwise the process of new bone-mass generation would not start. In the present case we find that $C^*$ is at the level comparable to $B^*$ at small $p$ and steadily increases with increasing $p$. Based on these observations, the natural conclusions can be drawn as, rate of differentiation of blast cells from mesenchymal cells, if enhanced by any means, would, in turn, enhance considerably the rate of formation of bone-mass. So, increase of $p$ favours generation of new bone-mass.

Phase diagram in Figure 8 shows that $B^*$ and $S^*$ assume their respective global stable values with very moderate increase of proliferation rate of osteoblast by cytokines ($\alpha$), whereas osteoclast population $C^*$ increases linearly at a very fast pace with $\alpha$. So, keeping the value of $\alpha$ at a suitable level below 0.5 would mean achieving a control over new bone-mass generation rate.

Results of calculations as depicted in Figure-9 signify that both $B^*$ and $S^*$ are diminished with the enhancement of the clast-cell generation-rate (through differentiation from monocytes or macrophages) and steady parabolic increase of $C^*$. Thus, enhancement of the numerical value of $q$ beyond its default level is observed not to be favourable for new bone-mass generation.

In Figure 10 we find that asymptotic stable values of osteoblast ($B^*$) and osteocytes ($S^*$) keep increasing linearly and steadily with the increase of $\mu$, the apoptosis rate of osteoclast cells. But the osteoclast stable value $C^*$ falls sharply at very small $\mu$ and beyond the value of $\mu{\sim}0.1$, $C^*$ becomes independent of $\mu$ assuming global stability at $C^*{\sim}300$. The situation here at moderate values of $\mu$ (apoptosis rate of osteoclast cells) becomes conducive for new bone-mass generation satisfying all required conditions.

Values of $B^*$ and $C^*$ does not at all respond to changing values of $\lambda$, the rate of blast-clast interaction giving into the osteocyte population, rather they remain at their respective constant values $B^*{\sim}490$ and $C^*{\sim}320$. But the value of $S^*$ increases steadily and at considerable pace starting from its zero value at $\lambda \to 0$. Again, the situation here is favourable for new bone-mass generation.

We have also checked the phase diagram viewgraphs in terms of parameters $\gamma$ (removal of blast owing to their interaction with clast), $\eta$ (production rate of clast from blast-clast interaction), $\nu$ (rate of bone resorption) and $\mu'$ (natural removal rate of osteocytes). In none of these cases conditions for new bone-mass generation, i.e. maintaining moderate values of blast and clast populations ($B^*$ & $C^*$) and high value of osteocyte population ($S^*$), has been

## VI. SUMMARY

To summaries our work, we have formulated a theoretical archetype of the dynamic process of bone-mass formation. The archetype has been judged through analytical tools to see through its sustenance and to gain confidence about the stability of its time series solutions at the asymptotic time scale. The archetype is also put to numerical simulation and the associated results are analysed in detail through planer viewgraphs to arrive at certain predictive conclusions.

We find that change in parameters $p$ (generation rate of osteoblasts), $\alpha$ (lysis-type proliferation of osteoblasts), $\mu$ (apoptosis rate of osteoclasts) and $\lambda$ (generation rate of osteocytes from blast-clast interaction) lead to conditions favourable for new bone-mass generation. We assert that tuning of these parameters to specific suitable levels would definitely produce new bone-matter of any desired magnitude. In other words, bone-matter generation can be controlled through changes of numerical values of parameters $p, \alpha, \mu \& \lambda$. This predictive conclusion is an outcome of our analysis of the theoretical archetype which can be tested in future clinical trials or in-vitro experimentation. This very concept would also be useful in future endeavour of framing any non-invasive mechanism of fracture healing.

In our future research efforts in the area of bone-formation dynamics, we would take up the problems justifying the numerical (default) values of various archetype-parameters and exploring biological or mechanical analogues of the mentioned favourable parameters ($p, \alpha, \mu \& \lambda$) through which tuning of these parameters could be achieved. We would also try to include electromagnetic field within the theoretical archetype with the focus to gather precise knowledge about non-invasive fracture healing in bones.

## Acknowledgements

One of the authors N. Hui would like to thank UGC, India for partial financial support in the form of a Minor Research Project No. PSW-69/12-13 (ERO).

**References**

*Corresponding Author: Biplab Chattopadhyay, E-mail: bchattopa@gmail.com

[1] D. J. Griffiths, *Introduction to Electrodynamics*, Pearson Education Limited, USA (2012).
[2] R. L. Duncan and C. H. Turner, Mechanotransduction and the functional response of bone to mechanical strains. Calcif. Tissue Int. 57 (1995) 344-358.
[3] S. C. Cowin, S. Weinbaum and Y. A. Zeng, A case for bone canalliculi as the anatomical site of strain generated potentials. *J. Biomech.* 28 (1995) 1281-1297.
[4] F. Shapiro, Bone development and its relation to fracture repair. The role of mesenchymal osteoblasts and surface osteoblasts. *Euro. Cell. Mater.*.15,(2008) 53-76.
[5] C. A. Basset and R. O. Becker, Generation of electric potentials by bone in response to mechanical stress. *Science.* 137 (1962) 1063-1064; C. A. Basset, R. J. Pawluk and R. O. Becker, Effects of electrical currents on bone in vivo. *Nature.* 204 (1964) 652-654; C. Andrew and L. Bassett, Beneficial Effects of Electromagnetic Fields. J. Cell. Biochem. 51 (1993) 387-393.
[6] W. Kuang and S. O. Nelson, Low-Frequency Dielectric Properties Of Biological Tissues: A Review with Some New Insights. Amer. Society. Agri. Eng. 41 (1998) 173-184.
[7] N. M. Shupak, Therapeutic Uses of Pulsed Magnetic- Field Exposure: A Review. Radio Sci. Bulletin. 307 (2003) 9-32.
[8] A. A. Pilla, Mechanism and therapeutic applications of time-varying and static magnetic fields, in: F. Burns and B. Greenebaum (Eds.), *Handbook of Biological Effects of Electromagnetic Fields*, 3$^{rd}$ Edition, Taylor & Francis, UK (2006), pp. 351-409.
[9] C. Zhong et al., Effects of electromagnetic fields on bone regeneration in experimental and clinical studies: a review of literature. Chinese Med. J. 125 (2012) 367-372.
[10] C.Delloye, et al., Bone regenerate formation in cortical bone during distraction lengthening: an experimental study. Clin. Orthop. Rel. Res. **250** (1990) 34-42.
[11] E Chao and N. Inoue, Biophysical stimulation of bone fracture repair,

regeneration and remodelling. *Eur. Cell Mater,* 6 (2003) 72-85.

[12] A. M. Perfitt, The cellular basis of bone turnover and bone loss. Clin. Orthop. Rel. Res. **127** (1977) 236-247.

[13] F. Forriol, F. Shapiro, Bone development. Interaction of molecular components and biophysical forces. *Clin. Orthop. Rel. Res*. 432 (2005) 14-33.

[14] P. J. Nijweide et al., *Principles of Bone Biology,* Edited by J. P. Bilezikien et al, Academic Press, New York (1996), P. 115.

[15] M. Zaidi et al, Osteoclast function and its control. Exp. Physiol. 78 (1993) 721-739.

[16] S. M. Sims and S. J. Dixon, Inwardly rectifying K$^+$ current in osteoclasts. *Amer.J.Physio*. 89 (1989) C1277-1282.

[17] B. A. Scheven, J. W. Visser and P. J. Nijweide, In vitro osteoclast generation from different bone marrow fractions, including a highly enriched haematopoietic stem cell population. Nature. 321 (1986) 79-81; P. J. Nijweide and R. J. Mulder, Identification of osteocytes in osteoblast-like cell cultures using a monoclonal antibody specifically directed against osteocytes. Histochem. 84 (1986) 342-347.

[18] C. Palumbo, S. Palazinni, G. Marotti, Morphological study of intercellular junctions during osteocyte differentiation. *Bone*. 11 (1990) 401-406.

[19] C. V. Gay, Mechanisms of bone resorption. *Rev. Poult. Biol*. 1, 197-210 (1988); C. V. Gay, V. R. Gilman and T. Sugiyama, Perspective on Osteoblast and Osteoclast Function. *Poultry Sci*. 79 (2000) 1005-1008.

[20] J. Graham, B. Ayati, P. Ramakrishnan, J. Martin, Towards a new spatial representation of bone remodeling. *Math. Biosci. Engg.* 9 (2012) 281–295; J. Graham, B. Ayati, S. Holstein, J. Martin, The Role of Osteocytes in Targeted Bone Remodeling: A Mathematical Model. *PLoS One* 8 (2013) e63884.

[21] R. Barron, Anatom. Moleculer mechanisms of bone resorption by the osteoclast. *Anatom. Rec.* 224 (1989) 317-324.

[22] S. L. Teitelbaum, Y. Abu-Amer and F. P. Ross, Moleculer mechanisms of bone resorption. *J. Cell. Biochem*. 59 (1995) 1-10.

[23] A. Van der Plass and P. J. Nilweide, Isolation and purification of osteocytes. J. Bone Miner. Res. 7 (1992) 389-396; A. Van der Plass et al., Characteristics and properties of osteocytes in culture, *J. Bone Min. Res.* 9 (1994) 1697-1704.

[24] T. A. Franz-Odendaal, B. K. Hall and P. E. Witten, Buried alive: how osteoblasts become osteocytes. Dev. Dyn. 235 (2006) 176-190.

[25] A. Boyde, Metab. Evidence against "osteocytic osteolysis." *Metab. Bone Dis. Rel. Res.* 2S (1980) 239-255.

[26] E. H. Burger et al., Functions of osteocytes in Bone-Their Role in Mechanotransduction, *J. Nutri.* 125 (1995) 2020S-2023S.

[27] T. J. Heino, T. A. Hentunen and H. K. Vaananen, Conditioned medium from osteocytes stimulates the proliferation of bone marrow mesenchymalstem cells and their differentiation into osteoblasts. *Exp. Cell. Res.* 294 (2004) 458-468.

[28] T. C. A. Phan, J. Xu and M. H. Zheng, Interaction between osteoblast and osteoclast: impact in bone disease. *Histol Histopathol,* 19 (2004) 1325-1344.

[29] T. J. Chambers, Osteoblasts release osteoclasts from calcitonin induced quiescence. *J Cell. Sci*. 57 (1982) 247-260.

[30] B. M. Thomson, G. R. Mundy and T. J. Chambers, Tumor necrosis factors alpha and beta induce osteoblastic cells to stimulate osteoclastic bone resorption. *J. Immunol.* 138 (1987) 775-779.

[31] T. J. Chambers, The pathbiology of osteoclast. *J. Clin. Patho.* 38 (1985) 241-252.

[32] S. Zhao et al., MLO-Y4 osteocyte like cells support osteoclast formation and activation. *J. Bone Miner. Res.* 17 (2002) 2068-2079.

[33] S. D. Tan et al., Osteocytes subjected to fluid flow inhibit osteoclast formation and bone resorption. *Bone.* 41 (2007) 745-751.

[34] A. Bakker, J. Klein-Nulend and E. Burger, Shear stress inhibits while disuse promotes osteocyte apoptosis. *Biochem. Biophys. Res. Commun.* 320 (2004) 1163–1168.

[35] B. Chattopadhyay and N. Hui, Immunopathogenesis in Psoriasis through a Density-type Mathematical Model. *WSEAS Trans. on Math*. 11 (2012) 440-450.

[36] M. Cot, *Elements of Mathematical Ecology*, Cambridge University Press, USA, 2001.

[37] W. H. Press, B. P. Flannery, S. A. Teukolsky and W. T. Vetterling, *Numerical Recipes*, Cambridge University Press, USA, 1989.